# Goal oriented indicators for food systems based on FAIR data

Ronit Purian

Tel Aviv University

purianro@tauex.tau.ac.il

**Abstract**

Throughout the food supply chain, between production, transportation, packaging, and green employment, a plethora of indicators cover the environmental footprint and resource use. By defining and tracking the more inefficient practices of the food supply chain and their effects, we can better understand how to improve agricultural performance, track nutrition values, and focus on the reduction of a major risk to the environment while contributing to food security. Our aim is to propose a framework for a food supply chain, devoted to the vision of zero waste and zero emissions, and at the same time, fulfilling the broad commitment on inclusive green economy within the climate action. To set the groundwork for a smart city solution which achieves this vision, main indicators and evaluation frameworks are introduced, followed by the drill down into most crucial problems, both globally and locally, in a case study in north Italy. Methane is on the rise in the climate agenda, and specifically in Italy emission mitigation is difficult to achieve in the farming sector. Accordingly, going from the generic frameworks towards a federation deployment, we provide the reasoning for a cost-effective use case in the domain of food, to create a valuable digital twin. A Bayesian approach to assess use cases and select preferred scenarios is proposed, realizing the potential of the digital twin flexibility with FAIR data, while understanding and acting to achieve environmental and social goals, i.e., coping uncertainties, and combining green employment and food security. The proposed framework can be adjusted to organizational, financial, and political considerations in different locations worldwide, rethinking the *value of information* in the context of *FAIR data* in digital twins.

# Goal oriented indicators for food systems based on FAIR data

**Introduction**
There is currently a dire need in the improvement of both the quality and the quantity of food. Thus, a change in our food supply systems – and specifically agriculture – is clearly required. Agricultural research is a vital part of this transformation, and its development is key in understanding and dealing with the many problems in the current state of agriculture, as well as the environmental implications of food systems, and the interrelations with food security and the social practices of food production and consumption. Big earth data, socioeconomic surveys, demographic data, and many other sources for timely data should therefore be coordinated and integrated. This is a complicated endeavor where, in addition to the multidisciplinary nature of such research areas, efforts to encourage open science across academic research, industry research, and governmental studies for better operation and policy-making, increase the complexity.

Recently, the concept "digital twin" and its implementation in smart cities has been of great interest in various sectors. Digital twins are often envisioned as a means of collecting mass data for general-purpose use. However, creating specific goal-oriented indicators will provide much cleaner data, and provide a dimension through which to organize data, information, and knowledge for the relevant issues.

For many urban questions, new data sources with greater spatial and temporal resolution are required (OECD, 2019; OECD.Stat, n.d). When applying the food-supply use-case in an actual context, our focus is on the value of information that a digital twin may contribute to the environment, to people's health and to life quality in the city and in the region. In its first stage, this project introduced a highly advanced solution, based on hyperlocal data, to select a cost-effective location for urban farms. Detailed indicators were under development, considering existing data for reliable comparisons. Moreover, practices along the food supply chain, in several value cycles locally and globally – and with relation to quality standards, regulation, and new legislation – set a clear list of goal-oriented indicators.

This top-down approach is complementary to a bottom-up examination of data. Heat waves and water scarcity, green gas emissions, and food insecurity, are tremendesly disruptive to environmental, social, and economic systems. These of course include food systems, from production (e.g., an economic threat to the agricultural sector) to consumption (e.g., food scarcity). Such risks and uncertainties change the way we choose to conceptualize the digital representation. Developing the case study in Italy emphasized the role of local conditions and considerations in the design of valuable information systems.

To lay the foundation for a feasible and effective data system, first we wish to outline some possible indicators, rather the providing an extensive list of measures. To support decisions towards high-priority actions, we describe the vulnerability in a region and provide the reasoning for a cost-effective use case in the domain of food. This is a delicate process requiring attention to needs, identification of market anomalies, and other considerations. Overall, to create a valuable digital twin, we wish to outline the rationale towards significant adaptations – not necessarily a major change – by flexibly adjusting to a more viable way of function and structure in the city and in a region.

**The project**
The potential benefit of urban and peri-urban agriculture (UPA) to the environment is a main topic in sustainability research, followed by economic and social pillars. When comparing the thematic outcomes in the literature on sustainable and healthy cities, subjective wellbeing and food and nutritional security are leading themes. However, knowledge gaps still exist and create barriers in cultivating sustainability and wellbeing through urban policy and planning (Nitya et al., 2022). We wish to focus on the design of urban use-cases for digital twins, and help close these gaps.

First, we will illustrate a comprehensive enabling framework for monitoring and evaluating resource productivity and waste, agriculture's environmental performance, and the economic and social interrelations.

A coordinated effort is needed to ensure the development of valid evaluation criteria. Since our intention is to also proceed, in the near future, into a broad exploration and consolidation of ecosystem goods and services, the indicators were divided into three sections of objectives, as follows:

(1) **Environmental** section, emphasizing *footprint indicators*, with the objective to transform use of materials and waste generation throughout the food production-to-consumption processes.

(2) **Socioeconomic** section, devoted to *inclusive green economy*, with the objective to increase eco-entrepreneurship, food security and social resilience.

(3) **Data strategy** section, enabling *knowledge sharing*, with the objective to introduce a comprehensive framework for smart, green, and fair urban ecosystems.

Through the very basic production and consumption practices of food, we provide a solid framework for resource efficiency, circular economy, and the transition towards green recovery and growth and community building. This work is summarized in Table 1.

*Table 1. Green transition and food security through urban farming and hyperlocal climate sensor data*

| Domain | Environmental section | Socioeconomic section | Data strategy section |
|---|---|---|---|
| Focus | **Footprint indicators** > Methane emissions, water | **Inclusive green economy** > Job security and skills > in the Agricultural sector | **FAIR implementation** > Human & machine readability |
| Goal | To transform use and waste generation along the food production-to-consumption processes, e.g., to improve air quality in city neighborhoods. | To increase food security, public health, and social resilience, e.g., to provide fresh healthy food grown within the city. | To introduce a comprehensive framework for smart, green, and fair urban ecosystems, e.g., to increase resilience to climate change and to social risks. |
| Topic and expected outcomes | Emissions from road transport for food delivery: to reduce emissions of greenhouse gases and fossil fuels. Agriculture's environmental performance: to reduce soil loss in agriculture; to reduce energy demand in buildings through green roofs. Waste generation: to reduce the consumption of packaging, plastic materials, and water. | Local economy, food security, and health: to encourage public participation and education for all; to provide the infrastructure for a new ecosystem of local food markets; to promote collaborations and new business models. Green employment: to support nutrition and environmental training; preparing for aging population and green job creation. | Stakeholder engagement: to share knowledge and propose principles for FAIR data governance; to propose local and global collaborations and partnership models. Enabling framework: to demonstrate the value of urban agriculture and local consumption as a holistic enabling framework towards green transition. |

How fine is the resolution we need? We will proceed from the wide range of data coverage to the most cost-effective data specification.

**Environmental indicators – footprint and technologies**
Our solution comprises the many stages that food products go through, from production to consumption, and by that we take stock of current consumption of plastic materials, use and waste generation, emissions from road transport for food delivery – compared to zero-packaging and freight trucks, and minimizing water consumption (domestic, tourism, etc.) in a local agriculture market with direct producer-to-consumer coordination.

To map the environmental impact on a global level we wish to uncover the underlying effects of current systems, e.g., soil losses in agriculture, while promoting eco-entrepreneurship.

Green indicators, such as UNEP Indicators for Green Economy, include many other *footprint indicators*, from production to consumption. For example:

- Emissions of transport (cut by X% in Y years)
- Soil losses in agriculture

- Waste disposal by hotels and restaurants (tonnes/year). Example: "Total waste in Mauritius amounts to 416,000 tonnes of solid waste in 2009 (2011)".
- Energy and water consumption in hotels and restaurants (ktoe and m3 /year). Example: "Water consumption from domestic, industrial and tourism accounts for 205 m3 /year or 27% of total water used (2012)".

Indicators were collected and curated for a pilot project in Genoa, a port city in the Italian region of Liguria, and the sixth-largest city in Italy. In Genoa, there is a unique link between tradition and innovation (Riahi et al., 2017). The data collectively generated in this project holds promise to begin to integrate computational models associated with multiple urban sectors and data sources, including a potential for fulfilling the promise of crowdsourced data.

At this stage, *footprint indicators* and *inclusive green economy* are reviewed to map existing evaluation frameworks for different goals, ecosystems, and stakeholders, including companies, central and local governments, and cities. This stage provides the **Generic Frameworks**. Based on the current knowledge about footprint indicators and inclusive green economy, we will establish the environmental and socioeconomic pillars – on which to prepare a **Federation Deployment**. This prospective stage will be facing reality, i.e., the local factors (including conditions of demand and supply, and even political forces) that determine the feasibility of any model and technological solution. Barriers are unavoidable, however, the local constraints are not necessarily an obstacle. In the process proposed here – moving from the holistic view of generic systems to local needs – local barriers turned out to be main features for the design and implementation of a digital twin solution, and intrinsic success factors.

Goals and measures – as mentioned above – aim at representing ecosystems. The evaluation frameworks, presented in Tables 2-3, address the need in a holistic view; each outlines an evaluation perspective and the reasoning that account for certain domains and stakeholders.

*Table 2. Evaluation frameworks for governments, companies, and cities*

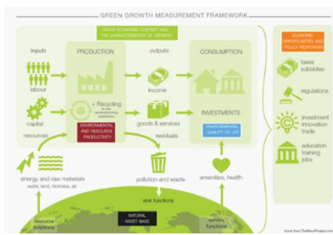

Green growth (OECD, 2017)

From environment and green *growth* (2017) to environment and post-Covid green *recovery* (2022) policies and measures

**For governments** to reduce emissions of greenhouse gases and the share of energy demand met with fossil fuels, and increase energy security; to align the spread of investment across policies and key sectors such as agriculture, waste management and forestry; to invest more in skills and innovation.

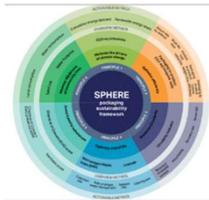

SPHERE (WBCSD, 2022)

Sustainability-waste-packages

**For companies** to monitor and minimize their packaging's contribution to climate change and nature loss.

**Principles** include packaging efficiency circularity, impact on climate change and biodiversity loss, absence of harmful substances and waste mismanagement.

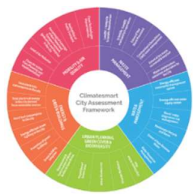

Cities Assessment Framework 2.0 (ClimateSmart, 2021)

Climate readiness and development

**For cities** while planning and undertaking development projects, towards a resilient urban habitat.

**Principles** include urban planning, green cover and biodiversity; energy and green buildings; mobility and air quality; water management; and waste management.

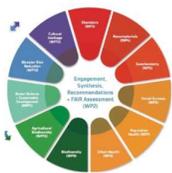

WorldFAIR (Codata, 2022)

Global cooperation on FAIR data policy and practice

**For reseachers and policymakers** to adapt FAIR Implementation Profiles to each (cross-)discipline area.

**Principles** include Cross-Domain Interoperability Framework with 11 case studies from the physical, agricultural and environmental sciences (including chemistry, nanomaterials, geochemistry, ocean data, disaster risk reduction), the social sciences (social surveys data; population health surveys with clinical and genomics data for COVID-19 research in eastern and southern Africa), urban health, biodiversity (digital extended specimen), and the cultural heritage sector (digital representation of heritage artefacts).

*Table 3. The role of climate indicators and sustainable development goals (SDGs) in evaluation frameworks*

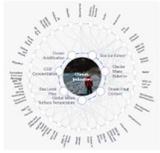 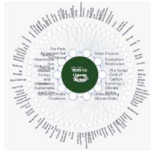

Sustainable development goal 13: Climate Action (right) and Climate Indicators Curation (left), World Meteorological Organization (WMO) are among the examples provided by the WEF (n.d.) for a holistic approach.

The indicators imply the problems a system aims to address.

At this stage after obtaining more information about the available data, the main attempt was to incorporate environmental goals into indicators e.g., to reduce emissions of greenhouse gases and fossil fuels.

Hyperlocal resolution data to support decisions such as the location of urban farms, creating a new urban habitat; or implementing renewable energy sources, lighting infrastructure, and more – were linked to goals such as, to reduce energy demand; to increase energy security. Mobile sensors were examined to collect high-resolution spatiotemporal data and carefully create, with unique algorithms, valuable climate information.

To propose a valid agenda to the digital twin industry, however, and to support its viability, the local factors must be considered with high attention to essential details. Rather than a detailed review of current indicators, a holistic yet selective view is proposed to select factors of high impact and to inspect their specific influences and possible interactions.

Encapsulating methane data should be a major goal in the construction of fine-grained data systems. Methane is a layer on which we present the building of the system (Why methane), further reviewing the state of agricultural methane emissions (Why food); and farming sector weaknesses and threats, strengths and opportunities (Why Italy).

In the conflicted playground of the climate agenda, methane plays the role of a tiebreaker. If we want to lead change, the sandbox of climate technologies must identify methane as a game changer.

## Why methane

Methane is on the rise in the climate agenda and when considering the benefits and costs of methane mitigation, it is evident that policies and tools to achieve this goal should be of priority in the multifaceted domain of food systems (e.g., Crippa et al., 2021; IEA, 2022; Qu et al., 2021; Shepon et al., 2018; Tepper et al., 2021).

Methane reduction is expected to have "greater climate benefits", and its increase is expected to have "greater adverse climate impacts", according to the recent IPCC report on "Emissions trends and drivers" (IPCC, 2022; chapter 2, p. 17). From the mitigation perspective, the adverse results – as well as the opportunity for improvement – make methane a key element in the construction of a valuable set of indicators.

The reduction of methane is essential and, to cope with the global warming, a rapid and sustained reduction is also achievable (IEA, 2022). This leads to the question of how much can be mitigated, and by whom – which sectors can have a meaningful impact?

## Why food

Agriculture is responsible for most methane emitted into the atmosphere, more than the energy sector which accounts for about "40% of total methane emissions attributable to human activity", according to Global Methane Tracker (IEA, 2022) – even when considering the increase in energy-related methane emissions as of post-pandemic higher demand and production of fossil fuel. Still, data is difficult to achieve. Global efforts to track methane emissions must rely on estimations, measurement initiatives, and scientific research, in addition to data reported by official public bodies to the UN Framework Convention on Climate Change (UNFCCC), which are "not yet accurate enough" (IEA, 2022), as can be seen in Figure 1.

*Figure 1. Global methane emissions by sector as of February 22, 2022: Estimates (light blue) higher than official reports (blue)*

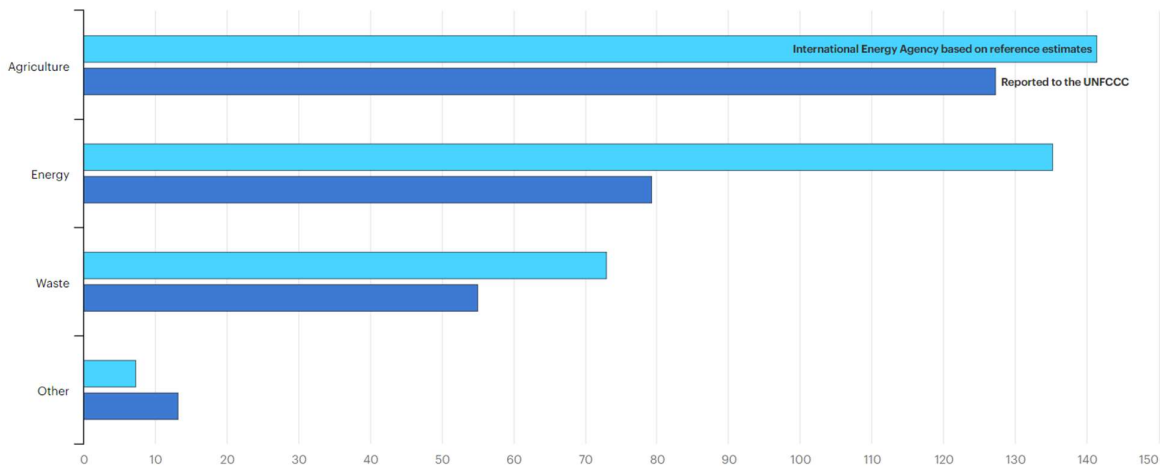

IEA (light blue): International Energy Agency (IEA) based on data measured in scientific studies, measurement initiatives and estimations

UNFCCC (blue): UN Framework Convention on Climate Change (UNFCCC) based on data reported by official public bodies

Studies persistently show the strong connection between methane and food systems (Porter, 2022). Recent studies, backed by new global databases, meta-analysis and systematic reviews, help understanding the numbers and options for abatement (Almeida et al., 2021; Crippa et el., 2021; Dorr et al., 2021a; 2021b). Crippa et el. (2021) developed a food emissions database (EDGAR-FOOD), based on the Emissions Database of Global Atmospheric Research (EDGAR), and the FAOSTAT database for land use and related emissions data. Land use change is a major factor in the estimations of food emissions when accounting for the alternative option of green land use. Altogether, agriculture and land use activities are responsible for 71% of global anthropogenic GHG emissions, with the remaining resulting from the various activities along the supply chain of food, including transport and fuel production, packaging and retail, consumption and waste management.

### Why Italy
A global look at methane emissions reveals that, despite an overall rise worldwide, emissions in the agricultural sectors are being reduced. The effort to reduce methane emissions has been successful in Italy, on a national level. However, the reduction occurred despite an increase in the agricultural sectors in Italy.

Figure 2 shows the worldwide increase in methane emissions (a) and decrease in recent years in the agricultural sector (b). Conversely, Italy performed well in general emission reduction (c), but an increase is shown in agricultural methane emissions (d). To realize the salient inversion, we also show the % of change (e-f).

*Figure 2. Methane emissions worldwide and in Italy*

| | **Methane emissions** (kt of CO2 equivalent) | **Agricultural methane emissions** (thousand metric tons of CO2 equivalent) |
|---|---|---|
| Worldwide | Figure a. Source: https://data.worldbank.org/indicator/EN.ATM.METH.KT.CE | Figure b. Source: https://data.worldbank.org/indicator/EN.ATM.METH.AG.KT.CE |
| Italy | Figure c. Source: https://data.worldbank.org/indicator/EN.ATM.METH.KT.CE?locations=IT | Figure d. Source: https://data.worldbank.org/indicator/EN.ATM.METH.AG.KT.CE?locations=IT |
| | Data: World Bank; Climate Watch DATA, GHG Emissions. World Resources Institute, Washington, DC. https://climatewatchdata.org/ghg-emissions | |
| | Methane emissions **(% change from 1990)** | Agricultural methane emissions **(% of total)** |
| | Figure e. Source: https://data.worldbank.org/indicator/EN.ATM.METH.ZG?locations=IT | Figure f. Source: https://data.worldbank.org/indicator/EN.ATM.METH.AG.ZS?locations=IT |
| | Data: World Bank; European Commission, Joint Research Centre (JRC)/ Netherlands Environmental Assessment Agency (PBL); Emission Database for Global Atmospheric Research (EDGAR) https://edgar.jrc.ec.europa.eu | |

National strategies for reducing on-farm methane emissions, however, often raise doubts and uncertainties. Representative organizations of the farmers resist their inclusion in the Emissions Trading Scheme (ETS) as governmental incentives are often conditioned with performance measures and possible taxes (e.g., FAO's Livestock, Climate and Environment community-of-action with the IPCC; He Waka Eke Noa in New Zealand, and more).

The societal aspects along the food supply chain cover a broad scope of life domains (O'Neill at el., 2017). Before presenting the broad view of societal issues, and in consistence with the process of eliminating the environmental indicators, a focus is proposed on a sector in crisis, the farming sector in Italy.

### Climate events and agriculture in Italy

Drought, heat waves, and poor technical infrastructure put the agricultural sector in Italy at risk. A third of the national agricultural production originates from the farms in the area of the Po River, and the prolonged drought has dried up the river. The government has declared a state of emergency, with the forecast for enhanced decrease in available water resources, expecting a decrease of up to 40% (AP, 2022).

The drought and the lack of effective water infrastructure, in the Northern regions in particular, has put at risk half of the livestock national production, according to a special brief by EURACTIV Network (2022) that tracks the effects of climate change on the farming sector, and "how famers are coping with living life on the edge". A coverage by the Washington Post (2022) on the drought in Italy provided a map of the Evaporative Stress Index (ESI) to illustrate the average amount of water evaporating from land surface and vegetation between June and July. Comparing with normal values, they concluded, "Basically, there is no water left".

The vulnerability of food production systems to climate change, and specifically the combination of poor water infrastructure together with drought, make the adoption of green technologies and skills an urgent need. However, new skills and green innovation that combines new technologies for agriculture are not easy to achieve, rather, insufficient assessment of training and innovation in green technologies represents a possible inadequacy in the measurement of both the green transition and its employment effects of economic recovery (OECD, 2022; Table 4).

### Green employment in Italy

Green jobs require skills transformation. As stated in recent reports on this timely topic, "Measuring the resulting employment effects of the green transition is challenging because it requires simultaneously evaluating all policies introduced and their direct and indirect economic and labour market impacts and interactions" (Cedefop, 2021; p. 16. Emphasis added). Moreover, the painful transition from *green growth* to post-Covid *green recovery* struggles in the area where it is most needed – in response to the economic breakdown, and sectoral unemployment, in different regions and territories.

When assessing the environmental impact of measures in the OECD Green Recovery Database (OECD, 2022, April 21; Table 1), green recovery fails to provide positive indicators (It provides a few positive ones, and no negative ones) in green skills transformation: "Relatively few recovery measures focus on skills training and on innovation in green technologies. This also represents a missed opportunity, as more attention to measures that can drive green job creation, notably to compensate for job losses in other industries, can help to ensure a 'just transition'" (OECD, 2022, April 19; 21. Emphasis added).

Table 4 shows the absence of "Skills training" in the OECD Green Recovery Database (which leads to the next section).

*Table 4. "Skills training" in the OECD Green Recovery Database, and (below) spread of investments across sectors in Italy*

| POSITIVE | Energy | Aviation | Ground transport | Maritime transport | Industry | Buildings | Agriculture | Forestry | Waste management | Other or Multiple | TOTAL |
|---|---|---|---|---|---|---|---|---|---|---|---|
| Tax reduction / other subsidy | 55 | 1 | 67 | 0 | 4 | 28 | 5 | 10 | 5 | 31 | 206 |
| Grant/Loan (including interest-free loans) | 99 | 2 | 110 | 7 | 23 | 72 | 21 | 13 | 12 | 96 | 455 |
| R&D subsidies | 32 | 7 | 15 | 1 | 5 | 7 | 1 | 1 | 2 | 37 | 108 |
| Regulatory change | 48 | 0 | 22 | 1 | 5 | 7 | 14 | 3 | 8 | 39 | 147 |
| Skills training | 1 | 0 | 0 | 0 | 2 | 0 | 0 | 0 | 0 | 27 | 30 |
| Other or not specified | 41 | 1 | 40 | 4 | 4 | 6 | 22 | 8 | 7 | 74 | 207 |
| TOTAL | 276 | 11 | 254 | 13 | 43 | 120 | 63 | 35 | 34 | 304 | 1153 |
| **NEGATIVE** | Energy | Aviation | Ground transport | Maritime transport | Industry | Buildings | Agriculture | Forestry | Waste management | Other or Multiple | |
| Tax reduction / other subsidy | 31 | 7 | 9 | 0 | 0 | 0 | 3 | 0 | 1 | 6 | 57 |
| Grant/Loan (including interest-free loans) | 9 | 35 | 6 | 1 | 1 | 0 | 0 | 0 | 0 | 4 | 56 |
| R&D subsidies | 1 | 0 | 0 | 0 | 0 | 0 | 0 | 0 | 0 | 0 | 1 |
| Regulatory change | 11 | 0 | 4 | 0 | 9 | 0 | 1 | 0 | 0 | 14 | 39 |
| Skills training | 0 | 0 | 0 | 0 | 0 | 0 | 0 | 0 | 0 | 0 | 0 |
| Other or not specified | 17 | 9 | 15 | 0 | 0 | 0 | 0 | 0 | 0 | 4 | 45 |
| TOTAL | 69 | 51 | 34 | 1 | 10 | 0 | 4 | 0 | 1 | 28 | 198 |

Source: Green recovery (OECD, 2022)

To conclude, the objective of green job creation through green initiatives, according to recent reports, is difficult to achieve.

When considering risk evaluation and characterization for adaptation and mitigation strategies, the agriculture crisis is both a threat (risk of losing third of Italy's agricultural production) and an opportunity – to helping the farming sector implementing new technologies; utilizing the market structure towards lean adaptation of small farms to a new mode of agile agriculture, and reducing anthropogenic greenhouse gas (GHG) emissions that contribute to climate change.

The purpose of indicators is, therefore, not only to provide a detailed account of what has been achieved. Choosing indicators is a process of narrating a story (Riahi et al., 2017).

Socioeconomic status of agricultural workers, specifically during times of low yield, is a promising axis on which to propose evaluation indicators. Thus, we propose to focus on the response needed in the farming sector when carrying out the objective of green job creation through green initiatives.

At the same time, this objective deserves a broader attention, beyond a state aid scheme for farmer, and towards rethinking urban nature and the many influences of green areas on human health and wellbeing.

A new and inclusive framework (especially for socioeconomic indicators) must consider, along the environmental aspects, the factors that affect and change our behaviors and urban lifestyle. What are the societal influences of technological acceleration, the service economy, and globalization on our life in cities? How does the sharing economy, in its current form, shape our service markets – and the habitat?

**Socioeconomic indicators – inclusive green economy**
The socioeconomic impact of food supply systems goes much beyond food security for all, in quality and quantity.

Socioeconomic implications include, among other things, preparing for aging population, e.g., through vocational training towards green jobs, health education to encourage nutrition-related behaviors, and other measures of health and wellbeing. Thus, the same questions presented in the **Environmental section** – how to choose **footprint indicators** and how fine is the resolution we need – are presented in the **Socioeconomic section** – how to proceed from the broad challenges of **inclusive green economy** towards effective intervention and evaluation. Who are the groups in focus, what are the goals (accounting for reciprocity, coping with uncertainties, green employment for green recovery – dedicated to vulnerable sectors, or specific social groups?), and how to measure (indicators that are part of the expected evaluation).

Chronic urban problems such as inequality, air pollution, traffic congestion and food insecurity affect us all. However, some groups are more vulnerable than others. Heat waves, flooding, droughts, and other effects of the climate crisis may leave them defenseless. Socioeconomic indicators and cultural values (surveys, public participation, etc.) support decisions in public projects. New climate risks to health and wellbeing, for different demographics, can be detected and recognized. However, while welfare is often associated with socioeconomic conditions, our intentions are to integrate the environmental aspects into a new framework of welfare, health, and wellbeing, to propose a valid agenda for digital twins as an emerging industry sector, and support its viability.

Size of cities, determined by population density, is often related to new life routines. The fast pace of life, globalization, and technological acceleration have caused new social and environmental problems (Purian, 2021; Ronen & Purian, 2021). In order to lead the movement forward towards fair and green smart cities, pragmatic principles and evaluation indicators are being applied, addressing climate threats. In line with the shared socioeconomic pathways proposed by O'Neill at el. (2017) and Riahi et al. (2017) for possible futures in the 21st century, current issues force us to focus on interrelations among various factors.

In this project we aim to draw attention to themes such as natural assets that organically coexists with the local history and other resources to preserve and cherish, in every city and region, in a unique urban lifestyle. To do that, we propose to combine indicators that are detailed along shared life domains.

**Market structure:** In different domains (e.g., MaaS, in Purian et al., 2019), attempts are made to achieve responsible models of a community-based sharing economy, and for technology design and adoption that considers coordination and fair access to services.

**Appreciation of nature:** Protecting urban biodiversity and the benefits of biodiversity for human well-being are related to food systems through appreciation of nature in urban agriculture, mainly edible trees, shrubs and bushes that are nutritious wild food.

**Urban cognition:** The current era of technological acceleration in today's global cities raises problems of stress, cognitive load, uncertainty, and a growing need for belonging. Urban cognition and sociological influences of the unprecedented scope of data we generate can be relieved in the presence of green areas. Even small green spots in the city may contribute for better physical and mental health.

**Social inclusion:** Key issues include social polarization, new types of digital divide and the crucial need in digital literacy (e.g., inoculation to fight misinformation), and other effects of size in many life systems. The implementation of new technologies, that generate enormous amounts of data, further accelerates these patterns and dynamics of social exclusion. While many cities cope with population boom, aging population, labor migration, and lack of working hands, we wish to capture an inclusive evaluation framework, considering the severity of disparities.

**Technological innovation for good:** Community gardens and urban farms that are equipped with facilities for nutrient management, integrated farming and precision agriculture systems are an antidote to the feeling of powerlessness, not only due to poverty, but also due to lack of influence and perceived control in the smart city.

Community building is related to the food domain in several ways, from community gardens to delivery, cooking and consumption, and surplus foods and waste management. As urban areas expand and more people live in cities, a growing environmental awareness is being developed. Resilience, the ability to affect main affectors, and to believe in their benevolence when facing large institutions, depends on improving environmental quality and natural hazard protection, as well as food security for all, in quantity and in quality.

Eco-entrepreneurship that combines the local economy, and encourages green employment and trade, has the advantages of circular economy while also contributing to social resilience.

### From a viewpoint of the consumers

Food security is an accurate manifestation of social and economic gaps, which do not necessarily coincide. The nutritional quality of food, health security, is almost a hidden aspect compared to the available supply in quantity. The ambiguous scope of food insecurity is core to challenges we face today. To what extent is food insecurity the result of "a lack of available financial resources for food at the household level" (proposed by Hunger + Health, 2022)? Rather than a household-level economic condition, a focus on community practices is proposed; extending merely financial disparities to indicate a broader view, namely, the social determinants of a poor diet that cause obesity, heart disease and many other medical conditions (e.g., FRAC, 2017).

### From a viewpoint of the producers

Uncertainties are often mentioned in the context of farming reforms to reduce agricultural emissions. Such uncertainties, however, create the organizing themes around which to establish the partnerships. The Partnership Platform (UN, 2022), for example, is a way to tackle uncertainties and to create the required knowledge networks. Rather than straitened reforms, national strategies for reducing on-farm methane emissions should be coordinating actions across sectors and facilitating the creation of ecosystems and collaborations that build trust.

### From a viewpoint of the digital twins

Threats and uncertainties are core to the design rationale of effective information systems, and digital twins are information systems that take issues of system goals and design to the extreme. The extremes are both in the need to integrate data from multiple sources, and in the need to address acute problems. Flexible design to address acute needs is possible when operating an infrastructure that requires openness, as a way of thinking. Openness is inherent to the system, from a holistic vision to data formats, throughout the data life cycle: in the format of FAIR data, which is necessary for machine readability as well as human understanding; and in the strategic vision that defines clear sets of goal-oriented indicators, organized along the main environmental, social, and economic dimensions of agriculture, nature, and food.

Openness is also consistent with the thrust of open science, and big earth data. After the problem statement, attaining a focused view of acute and complicated issues, next we develop the approach to address them with FAIR data that allow data partnerships and knowledge sharing, locally and globally.

**Data strategy – innovation in the physical twin, based on FAIR implementation in the digital twin**

Knowledge sharing receives a new meaning with the conceptualization of FAIR data and the actual implementation of data systems that are machine readable and human understandable. Considering the connection between methane and food systems, what should a digital twin solution introduce to our urban-data toolbox? Previous works proposed conceptual frameworks for designing and implementing digital twins. In this work we would like to take a step further towards implementation, into a structure of open and FAIR data (Schultes et al., 2022). This is one contribution of our work. The second contribution is with the harnessing of technology towards a resolution of crucial systemic problems. To do that, we should be narrating a meaningful story, the story of the food.

### WHAT: Use cases in food systems (the physical twin)

In the context of agricultural information systems, smart farming is an emerging domain (e.g., Verdouw et al., 2021). Beyond the agricultural aspects in knowledge-intensive farming, in this case study we demonstrate the need, and the ability, to integrate big and prominent issues, such as the methane question, into the design and actual operation of the physical farms and their digital twins. The new knowledge-rich agriculture is expected to be a source of data, shared among stakeholders (e.g., public participation in climate data – in local and global levels. Moreover, the intention when setting the focus on the food life cycle is to lay the foundation for new food systems – and ecosystems – that develop more sustainable work practices. To illustrate, possible use cases are presented, linked to main goals and questions along the design process:

### Smart farms to reduce food scarcity and environmental costs of meat consumption

We choose to apply a food systems perspective to climate change, mainly to reduce methane emissions from cattle. Reducing meat consumption may not include dairy cows, but many other ruminant livestock produce significant amounts of methane (e.g., sheep). To reduce meat consumption, vegetarian alternatives should be supplied. Mushrooms are nutritionally valuable as a substitute, and mushroom cultivation is common in knowledge-intensive agriculture and urban communities (e.g., the Fungi Academy https://fungiacademy.com). In addition, a systematic review and meta-analysis showed the advantage of local farms (Dorr et al., 2021a). Therefore, we propose to focus on urban mushroom farming (e.g., Dorr et al., 2021b).

This is a rather specific conclusion, reached based on the broad literature that already exists in this interdisciplinary topic. The current rich body of knowledge, in different domains, leads further on to more advanced decision points that require specific reference data for clear and measurable indicators.

Key decision points are primarily the use cases that design the digital twins, setting data requirements, as follows.

### Nutritious produce and food safety

Nutritional values of beef replacement plant-based diets, and specifically the nutritional value of mushrooms, are main targets to achieve when designing food systems that are environmentally optimal (i.e., not one at the expense of the other, but improving both, the quality of nutrition and of environmental performance). Based on the system inputs and outputs, tools for assessing sustainable healthy diets already exist, including the efficient use of energy, land, and sunlight (Eshel et al., 2016; FAO, 2022; Jackson at el., 2021; Leger et al., 2021; Smith et al., 2021; Tepper et al., 2021).

This leads to several use cases of nutritional value, e.g., branding products as highly nutritional; and coping with complicated challenges of food safety and restricted products (e.g., UPU, 2017).

### Dietary requirements for environmentally optimal supply

Moreover, dietary requirements can be a perspective through which to plan and better utilize the capacity of local agriculture, either on a national, regional or an urban level, according to household demand and environmental conditions, either current or predicted (Nixon & Ramaswami, 2022). Furthermore, to encourage change and facilitate the reduction of meat consumption, a recommended food basket should be developed, with adaptation to climate and economic conditions. The idea is to propose an innovative approach to public participation: setting the ultimate dietary recommendations – as a means of advancing nutritional perceptions, improving current themes about nutritional security from a health point of view. To satisfy this resolution, a digital twin depends on access to multiple data sources, and a capacity to integrate the interlinkages among the systems and ecosystems of climate, food, land, water and oceans, and people through direct and indirect links to the socio-economic system and human well-being, as presented by IPCC (2019) and FAO (2022).

### Standards to satisfy consumers, producers, and regulators

Local data should be globally available to improve national and international standards and regulations. Today, to comply with international standards on food quality (e.g., ISO), harvest may be delayed – to meet measures of size, weight, or other criteria. However, for some agricultural products, harvest may better be carried out earlier – to provide fresh crop more often and satisfy the consumer. Variables to consider span from consumer behavior and preferences on the one hand to operation management on the other; mainly available cultivation area vs. delivery costs, packaging, and emission.

### Who are the farmers

Green recovery and the promise of inclusive green economy account for green jobs in the general public, e.g., for workers who can no longer perform physically demanding roles such as construction workers, commercial drivers, building cleaning, personal care, food service workers (Ross & Bateman, 2019). New skills are of great relevance to the low-wage workforce after certain age. Other case studies on urban farms and agriculture focus on education programs in schools, working with youth, or preservation of traditional knowledge. Indeed, various socioeconomic indicators populate the Generic Framework (Table 1), considering existing evaluation frameworks (Tables 2-3) in the first stage. In the next stage, a case study in Italy is developed towards a Federation Deployment. The agricultural market structure of small local farms near the city makes it essential to support the farming sector. As presented earlier, this sector is vulnerable to the climate crisis, facing not only droughts but also insufficient water infrastructure. While an agricultural partnership must include the farmers, and help them acquire technical skills to apply technological innovation in agriculture – they are not the exceptional mitigators but a strong link in a system that learns how to priorities problems and solutions in a cost-effective workplan that utilizes resources.

In addition to the "Environmental indicators – footprint and technologies", and in terms of "Socioeconomic indicators – inclusive green economy", the purpose of this section is to make the adjustments to the economic and the political conditions; making it feasible, in each and every ecosystem, to adopt financial means and to flourish, and for different stakeholders to join and engage. Training new urban farmers while generations of farmers, who were born on farms, are struggling is ineffective, negligent, and politically impossible. At the same time, their representative organizations should not act on their exclusive behalf.

To improve food systems, there is an urgent need for selecting and fine-tuning the very specific factors along the narrative of food, e.g., choosing farm location based on land use, energy use, and other trade-offs. A clear focus exists, reducing greenhouse gas emissions as the main game-changer, among the many other resulting indicators and their implications. Accordingly, Ministries of Agriculture may increase the share of agriculture in GDP and eventually act for the benefit of farmers – and the entire public.

To coordinate the production and consumption of nutritious, affordable, and environmentally optimal food yield, across sectors and stakeholders, in different countries, the integration of data from multiple food systems is an important demand that digital twins may fulfill, if designed openly according to the FAIR principles.

### HOW: Software architecture (the digital twin)

Open science is a concept of inclusive shared knowledge, and FAIR principles are at the heart of open science.

A distinction should be made between the value of information (Purian, 2011), i.e., to assess the contribution of an information system to operation management, and the value of data (Mons et al., 2011), i.e., to enable the implementation of FAIR data practices in new ways that are essential to create strategic value.

### The operational value of information

Environmental footprint and resource use of urban agriculture are studied in recent years and provide a rather consistent view of trade-offs, suggesting decision points along the life cycle of circular urban farms, e.g., regarding location (even in proximity to end-users). As a reference for operation management: Niles et al. (2017) provide a "Food System Intervention Worksheet" with "Logical Approach to Support Prioritization of Interventions" in high, medium, and low-income countries, for various aspects along the food supply chain (Appendix 2 in Niles et al., 2017; pp. 70-72), next to a thorough review of Potential Interventions (pp. 32-69).

### The strategic value of FAIR data

While the direct target – to mitigate near-term climate change and improve food security – is well established in the literature (e.g., Shindell et al., 2012), the connection of these crucial issues to the FAIR data approach and infrastructure (Mons, 2020) is new.

A way to illustrate the value of data is through a connection of utility functions: either individually; in their *original* classes; or in *new* data clusters. The options are presented in Figure 3 in the form of networks among individual data points, classes of data points, and clusters of data points from different classes. In the paper on the "Value of Information in the Network" (Purian, 2011), collective action is analyzed when environmental and social criteria emerge, in addition to pragmatic criteria of existing systems and routines. An analytical model is developed (Table 5), a weighted graph representing "how information gains its value in the network, from different perspectives" (p. 9). As opposed to network models that represent size, i.e., the volume communication, the "agenda that organizations would adopt depends on the information that flows in the network, between individuals, organizations, governments, and other institutions in the society" (p. 13). Building on the UN's (2020; p. 12) SDG scheme for multi-stakeholder partnerships, we present the idea of a common axis or narrative, on which to organize the activity.

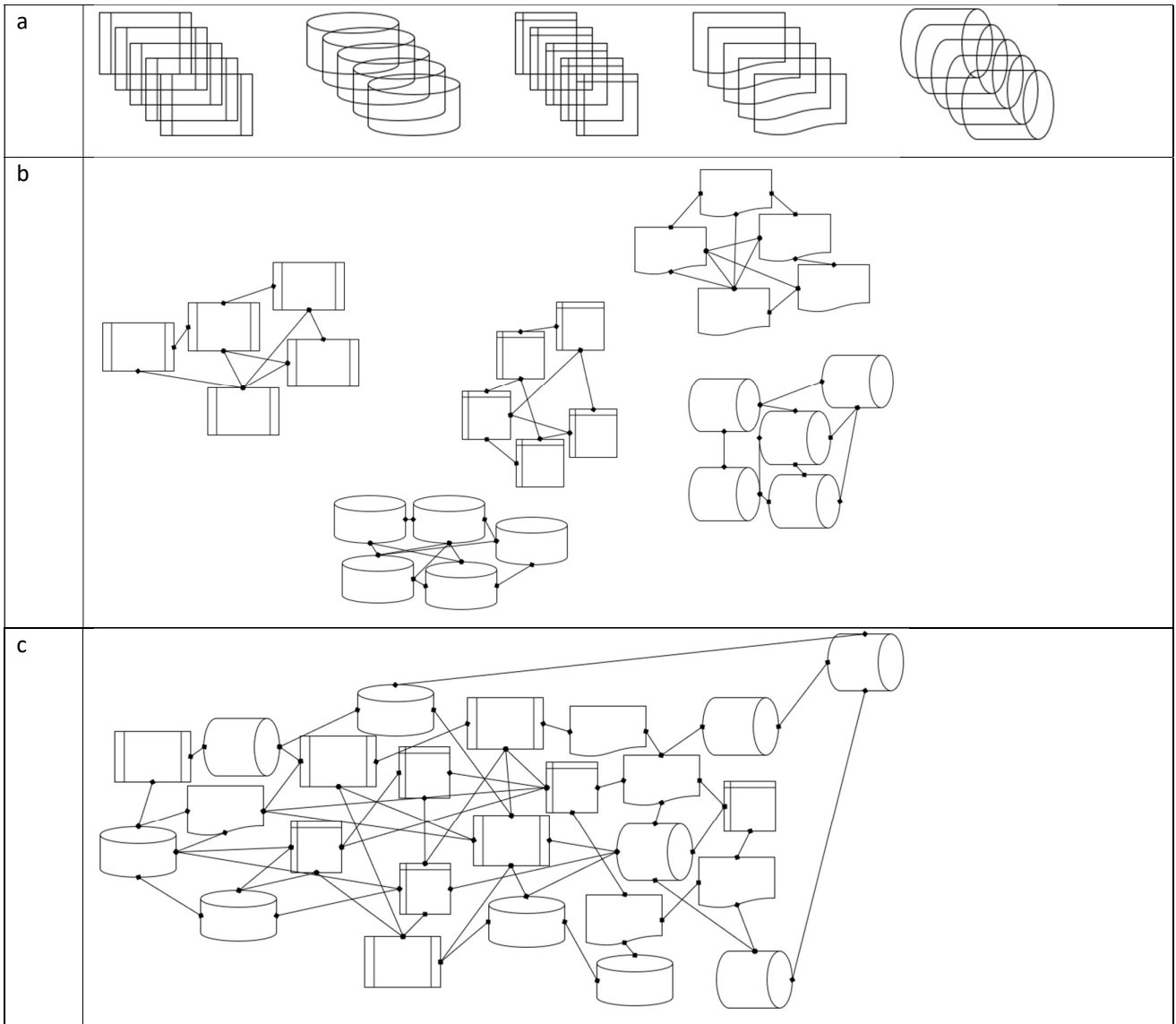

Figure 3. Individual data points (a), classes of data points (b), and clusters of data points from different classes (c) with a directed network

The directed network in Figure 3 c implies a directed acyclic graph (DAG) for Bayesian inference (Purian et al., 2022).

Utility functions, connected in networks across data sources, provide the value described in Table 5.

*Table 5. The value of information in the network vs. Reed's law, which describes the network size, as a reference*

| A network with: | Reed's law (size) | New model (value) | |
|---|---|---|---|
| One node value | 1 | $x_i \geq 0$ | $_i \in N$ |
| Connection | A | $y(x_i, x_j)$ | $_{i,j} \in E$ |
| Group | B | $z(x_i)$ | $_s \in P$ |
| Network of size N | $\alpha N + aN^2$ | $\alpha xN + yN^2$ | |
| Group-forming network (GFN) | $N + aN^2 + bZ^2$ | – | |

| | |
|---|---|
| The value $V$ of a network equals to the sum of value of node $i$ and the additional value from communication with node $j$ and the additional value from communication with a group of nodes $s$ as shown in the following equation: $$V = \sum_{i \in N} x_j + \sum_{i,j \in E} y_{ij} + \sum_{s \in P} z_s$$ Where: V = value in a network $i$ = identifies one node in the network $j$ = identifies one node in the network $s$ = identifies a group of nodes $x_i$ = the v of node $i$ $y_{ij}$ = the additional v from communication between node $i$ and node $j$. $z_s$ = the additional v from communication between a group of nodes $s$. | The questions the model describes, for each entity, are: $x_i$ – the value of the $i^{th}$ node (data source). How do we evaluate $x_i$? What influences it? $y_{ij}$ – the additional value from communication between the $i^{th}$ node and the $j^{th}$ node. How to measure $y_{ij}$? Is it none-negative? $z_s$ – the value added by a group of nodes. How to measure $z_s$? Is it none-negative? |

Source: Purian (2011)

While UN (2020) generally refer to "little collaboration" vs. "systematic collaboration" across sectors to create partnerships, our emphasis is on the integration of data (and indicators) across data sources.

One of the main questions, in the context of data management, is regarding the modularity level of data, and the resolution depends on the FAIR data life cycle along the proposed food supply chain. For this reason, implementing FAIR data operation is a strategic move. Life cycle assessments of novel agricultural technologies span across the many domains and externalities of food production and consumption. This requires detailed encapsulation of data.

Once the network is established, it matters whether the integration of data is creating a new axis of knowledge, or maintains existing routines. The meaning of data integration is to create new axes of action; a direction in Figure 3c.

To facilitate the coordination among stakeholders and sectors, and the creation of ecosystems and partnerships, FAIR data and open science for data-intensive research should be achieved with digital twins (Schultes et al., 2022).

Scenarios, data, and tools

**Scenarios:** To outline the implementation of open and FAIR data (Mons, 2020; Wilkinson et al., 2016) in digital twins, we need to set goal-oriented indicators. Choosing footprint indicators and inclusive green economy as pillars to the framework was the first stage (Generic Framework; to achieve this goal, we introduce some of the main indicators and evaluation frameworks that may be beneficial for the solution). Transition towards implementation, in the following stage (Federation Deployment; setting the groundwork through use cases from which to assemble the data needed), revealed the success factors that play a key role in each pillar. In the environmental pillar: When considering the influences of agriculture on the environment and vice versa, reduction of greenhouse gas emissions is key. In the socioeconomic pillar: When considering the need in skills and technological innovation for green recovery, it becomes evident that our emphasis is on a solution for agriculture and farmers. Accordingly, scenarios were proposed to innovate with the capabilities of digital twins.

**Data:** Our goal is to define an axis or a narrative through which to choose the appropriate indicators, thus making sure the indicators are goal-oriented, sharable, and machine-readable. The proposed narrative is the main axis on which to develop the ontology, and to make it possible to share the new knowledge among stakeholders: public participation in climate data, and a nourishing food basket that meets dietary requirements; evaluation frameworks and standards; and more, locally and globally. References are provided, throughout the paper, to resources for specific domain data.

**Tools:** Throughout the food supply chain, between production, transportation, packaging etc., a plethora of indicators cover the environmental footprint and resource use. However, tools and methods to organize the data and manage it are just starting to gain attention. Ontology-based access policies are among the excitatory approaches to utilize FAIR data, based on concepts and relations from a domain ontology, and providing licensing for both humans and machines (Brewster et al., 2020). Bayesian inference is applied in new ways, as an approach to detect bias in a dataset (Purian et al., 2022), and to provide an explicit characterization of data. Global emissions from livestock, for example, were examined with Bayesian inference to quantify methane emissions and attribute the largest anthropogenic source on the global scale (Qu et al., 2021).

Bayesian networks and ontologies are main data tools for the digital twin toolbox (Purian et al., 2022).

# Discussion

Creating **goal-oriented indicators** that enable coherent actions across environmental, social, and economic fields is challenging. To achieve this, we must provide a dimension – or a story – through which to organize data, information, and knowledge for the issues of wellbeing in cities, social resilience, and climate change, that are inherently linked. In the first stage of this study, we introduce some of the main indicators and evaluation frameworks, e.g., UNEP Indicators for Green Economy, that include many *footprint indicators*, from production to consumption (Tables 1-3). Improved consistency of geo-referenced high-resolution spatial data are among the major tools of the smart city. Accordingly, we examined a solution that combines urban farming and hyperlocal climate sensor data, towards food security and green growth. Circularity, a broader goal, also requires detailed information from multiple sources that may be beneficial for evaluation.

In this case study that examines the ability to incorporate a crucial, perhaps acute issue, into a long-term infrastructure of data techniques and methods. By doing that the contribution in twofold: choosing an urgent goal as an axis on which to organize a system; and going beyond a conceptual framework towards implementation.

Defining the question, setting the goal, is critical to the success of digital twins that aggregate enormous amounts of data. This prerequisite is shared by scientific disciplines and the common knowledge, from the very first map makers, to the golden age of information systems. Choosing just one example out of many, the coastline paradox (Mandelbrot, 1967) illustrates the need to set a scale that is relevant to the purpose of measurement; the length of the coast of England depends on the ruler's length. Since then, the field of information systems research insisted on the inner design of a data-information-knowledge continuum in systems (Simon, 1956). Setting the indicators means creating a representation of the city, therefore the importance of "narrating" the digital twin, re-establishing urban identity through technical decisions that may change the way the city acts and performs. This is the new science of management decision in the smart city.

# References

Almeida, A. K., Hegarty, R. S., & Cowie, A. (2021). Meta-analysis quantifying the potential of dietary additives and rumen modifiers for methane mitigation in ruminant production systems. Animal Nutrition, 7(4), 1219-1230. https://doi.org/10.1016/j.aninu.2021.09.005

AP (2022, July 13). Drought: Italy warns a third of farm production at risk https://apnews.com/article/agriculture-italy-climate-and-environment-77d11e4bd33ac53f4222ab66725cd68e

Brewster, C., Nouwt, B., Raaijmakers, S., & Verhoosel, J. (2020). Ontology-based access control for FAIR data. Data Intelligence, 2(1-2), 66-77. https://doi.org/10.1162/dint_a_00029

Cedefop (2021). The green employment and skills transformation: Insights from a European Green Deal skills forecast scenario. European Centre for the Development of Vocational Training (Cedefop), EU, Luxembourg http://data.europa.eu/doi/10.2801/112540

ClimateSmart Cities Assessment Framework 2.0 (2021, June). Cities Readiness Report https://niua.in/csc/assets/pdf/key-documents/Cities-Readiness-Report.pdf

Codata (2022). WorldFAIR: Global cooperation on FAIR data policy and practice https://codata.org/worldfair-global-cooperation-on-fair-data-policy-and-practice-a-major-two-year-project-starts-today; introduced https://www.scidatacon.org/IDW-2022/sessions/439

Crippa, M., Solazzo, E., Guizzardi, D., Monforti-Ferrario, F., Tubiello, F. N., & Leip, A. J. N. F. (2021). Food systems are responsible for a third of global anthropogenic GHG emissions. Nature Food, 2(3), 198-209. https://doi.org/10.1038/s43016-021-00225-9

Cusworth, D. H., Bloom, A. A., Ma, S., Miller, C. E., Bowman, K., Yin, Y., ... & Worden, J. R. (2021). A Bayesian framework for deriving sector-based methane emissions from top-down fluxes. Communications Earth & Environment, 2(1), 1-8. https://doi.org/10.1038/s43247-021-00312-6

Dorr, E., Goldstein, B., Horvath, A., Aubry, C., & Gabrielle, B. (2021a). Environmental impacts and resource use of urban agriculture: a systematic review and meta-analysis. Environmental Research Letters, 16(9), 093002. https://doi.org/10.1088/1748-9326/ac1a39

Dorr, E., Koegler, M., Gabrielle, B., & Aubry, C. (2021b). Life cycle assessment of a circular, urban mushroom farm. Journal of Cleaner Production, 288, 125668. https://doi.org/10.1016/j.jclepro.2020.125668

Eshel, G., Shepon, A., Noor, E., & Milo, R. (2016). Environmentally optimal, nutritionally aware beef replacement plant-based diets. Environmental science & technology, 50(15), 8164-8168. http://dx.doi.org/10.1021/acs.est.6b01006

EURACTIV Network (2022, Jun 17). Agrifood Special CAPitals Brief: Living life on the edge. https://www.euractiv.com/section/agriculture-food/news/agrifood-special-capitals-brief-living-life-on-the-edge

FAO (2022). Thinking about the future of food safety: A foresight report. Food and Agriculture Organization (FAO) of the United Nations. Rome https://doi.org/10.4060/cb8667en

FRAC (2017, December). The Impact of Poverty, Food Insecurity, and Poor Nutrition on Health and Well-Being. Food Research & Action Center (FRAC). Hunger & Health https://frac.org/wp-content/uploads/hunger-health-impact-poverty-food-insecurity-health-well-being.pdf

Hunger + Health (2022). What is Food Insecurity? Understand Food Insecurity. Hunger + Health and Feeding America https://hungerandhealth.feedingamerica.org/understand-food-insecurity

IEA (2022, February 22), Global Methane Tracker, International Energy Agency (IEA), Paris https://www.iea.org/reports/global-methane-tracker-2022; emissions by sector https://www.iea.org/data-and-statistics/charts/global-methane-emissions-by-sector-reported-to-the-unfccc-and-estimates-from-the-iea

IPCC (2022). Climate Change 2022: Mitigation of Climate Change. Working Group III, Intergovernmental Panel on Climate Change (IPCC) Sixth Assessment Report, UN https://www.ipcc.ch/report/ar6/wg3

IPCC (2019). Special Report on Climate Change and Land. Chapter 5: Food Security https://www.ipcc.ch/srccl/chapter/chapter-5

Jackson, R. B., Abernethy, S., Canadell, J. G., Cargnello, M., Davis, S. J., Féron, S., ... & Zickfeld, K. (2021). Atmospheric methane removal: a research agenda. Philosophical Transactions of the Royal Society A, 379(2210). https://doi.org/10.1098/rsta.2020.0454

Leger, D., Matassa, S., Noor, E., Shepon, A., Milo, R., & Bar-Even, A. (2021). Photovoltaic-driven microbial protein production can use land and sunlight more efficiently than conventional crops. Proceedings of the National Academy of Sciences, 118(26), e2015025118. https://doi.org/10.1073/pnas.2015025118

Mandelbrot, B. (1967). How long is the coast of Britain? Statistical self-similarity and fractional dimension. Science, 156(3775), 636-638. https://doi.org/10.1126/science.156.3775.636

Mons, B. (2020). The VODAN IN: support of a FAIR-based infrastructure for COVID-19. European Journal of Human Genetics, 28, 724–727. https://doi.org/10.1038/s41431-020-0635-7

Mons, B., van Haagen, H., Chichester, C. et al. (2011, March 29). The value of data. Nature Genetics. 43, 281-283. https://doi.org/10.1038/ng0411-281

Niles, M.T., Ahuja, R., Esquivel, J., Mango, N., Duncan, M., Heller, M., & Tirado, C. (2017, April 28). Climate change & food systems: Assessing impacts and opportunities. Meridian Institute, Washington, DC.

Nitya, R. A. O., Patil, S., Singh, C., Parama, R. O. Y., Pryor, C., Poonacha, P., & Genes, M. (2022). Cultivating Sustainable and Healthy Cities: A Systematic Literature Review of the Outcomes of Urban and Peri-urban Agriculture. Sustainable Cities and Society, 104063. https://doi.org/10.1016/j.scs.2022.104063

Nixon, P., & Ramaswami, A. (2022). County-level analysis of current local capacity of agriculture to meet household demand: a dietary requirements perspective. Environmental Research Letters, 17(4), 044070. https://doi.org/10.1088/1748-9326/ac5208

OECD (2022, April 21). Assessing environmental impact of measures in the OECD Green Recovery Database (Table 1, p. 8) https://www.oecd.org/coronavirus/policy-responses/assessing-environmental-impact-of-measures-in-the-oecd-green-recovery-database-3f7e2670

OECD (2022, April 19). The OECD Green Recovery Database: Examining the environmental implications of COVID-19 recovery policies https://www.oecd.org/coronavirus/en/themes/green-recovery

OECD (2020, February 27). OECD Agro-Food Productivity-Sustainability-Resilience Policy Framework: Revised Framework. Trade and Agriculture Directorate Committee for Agriculture. Working Party on Agricultural Policies and Markets https://one.oecd.org/document/TAD/CA/APM/WP(2019)25/FINAL/en/pdf

OECD (2019, February). Monitoring and evaluating agriculture's environmental performance https://www.oecd.org/agriculture/topics/agriculture-and-the-environment

OECD (2017). Green growth indicators. https://www.oecd.org/greengrowth/green-growth-indicators

OECD.Stat (n.d) Agri-Environmental Indicators (AEI) by countries https://stats.oecd.org//Index.aspx?QueryId=77296&lang=en

O'Neill, B. C., Kriegler, E., Ebi, K. L., Kemp-Benedict, E., Riahi, K., Rothman, D. S., ... & Solecki, W. (2017). The roads ahead: Narratives for shared socioeconomic pathways describing world futures in the 21st century. Global environmental change, 42, 169-180. https://doi.org/10.1016/j.gloenvcha.2015.01.004

Peh, H., Balmford, A., Bradbury, B., Brown, C., Butchart, H., ... & Birch, C. (2013). TESSA: A toolkit for rapid assessment of ecosystem services at sites of biodiversity conservation importance. Ecosystem Services, 5, 51-57.

Porter (2022). The question of Methane, by Shepon, A. On Borrowed Time: Campus discussion on climate change; in Planer Zero, Tel Aviv University Climate Crisis Initiative. https://www.sviva.net/wp-content/uploads/2022/02/מתאן-על-צהרים-דיון-the-quesiton-of-Methane-zoom-discussion.pdf

Purian, R. (2011). Value of Information in the Network: The Change in Pragmatic and Ethical Criteria. MCIS 2011 Proceedings. 107. https://aisel.aisnet.org/mcis2011/107

Purian, R. (2021). A Smart City Anomaly: The near becomes far, the far becomes near. GRF, Vol 40, No 1. https://grf.bgu.ac.il/index.php/GRF/article/view/596

Purian, R. and Ronen, O. (2021). Between Means and Ends – Sustainable and Smart Cities in Flux: An Editorial Introduction. GRF, Vol 40, No 1. https://grf.bgu.ac.il/index.php/GRF/article/view/606

Purian, R., van Hillegersberg, J. and Catlett, C. (2019, December). Life as a Service in the smart city: Fair play in information systems design, data integration and planning. ICIS, Munich https://icis2019.aisconferences.org

Purian R, Katz N, Feldman B, Ben-Yosef Y (2022) Unbiased AI. International Data Week (IDW) SciDataCon-IDW (AI & Reproducibility, Repeatability, and Replicability). https://www.scidatacon.org/IDW-2022/sessions/467/paper/1098

Qu, Z., Jacob, D. J., Shen, L., Lu, X., Zhang, Y., Scarpelli, T. R., ... & Delgado, A. L. (2021). Global distribution of methane emissions: a comparative inverse analysis of observations from the TROPOMI and GOSAT satellite instruments. Atmospheric Chemistry and Physics, 21(18), 14159-14175. https://doi.org/10.5194/acp-21-14159-2021

Riahi, K., Van Vuuren, D. P., Kriegler, E., Edmonds, J., O'neill, B. C., Fujimori, S., ... & Tavoni, M. (2017). The shared socioeconomic pathways and their energy, land use, and greenhouse gas emissions implications: an overview. Global environmental change, 42, 153-168. https://doi.org/10.1016/j.gloenvcha.2016.05.009

Ross, M., & Bateman, N. (2019, November 11). Meet the low-wage workforce. Metropolitan Policy Program at Brookings. http://hdl.handle.net/10919/97785; https://www.brookings.edu/wp-content/uploads/2019/11/201911_Brookings-Metro_low-wage-workforce_Ross-Bateman.pdf

Schultes, E., Roos, M., Bonino da Silva Santos, L. O., Guizzardi, G., Bouwman, J., Hankemeier, T., ... & Mons, B. (2022, May 11). FAIR Digital Twins for Data-Intensive Research. Frontiers in big Data, 5(42, 883341), 1-19. https://doi.org/10.3389/fdata.2022.883341

Shepon, A., Eshel, G., Noor, E., & Milo, R. (2018). The opportunity cost of animal based diets exceeds all food losses. Proceedings of the National Academy of Sciences, 115(15), 3804-3809. https://doi.org/10.1073/pnas.1713820115

Shindell, D., Kuylenstierna, J. C., Vignati, E., van Dingenen, R., Amann, M., Klimont, Z., ... & Fowler, D. (2012). Simultaneously mitigating near-term climate change and improving human health and food security. Science, 335(6065), 183-189. https://doi.org/10.1126/science.1210026

Simon, H. (1956). Rational choice and the structure of the environment. Psychological review, 63(2), 129. https://doi.org/10.1037/h0042769

Smith, P., Reay, D., & Smith, J. (2021). Agricultural methane emissions and the potential for mitigation. Philosophical Transactions of the Royal Society A, 379(2210), 20200451. https://doi.org/10.1098/rsta.2020.0451

Tepper, S., Geva, D., Shahar, D. R., Shepon, A., Mendelsohn, O., Golan, M., ... & Golan, R. (2021). The SHED Index: a tool for assessing a Sustainable HEalthy Diet. European Journal of Nutrition, 60(7), 3897-3909. https://doi.org/10.1007/s00394-021-02554-8

Tubiello, F. N., Rosenzweig, C., Conchedda, G., Karl, K., Gütschow, J., Xueyao, P., ... & Sandalow, D. (2021). Greenhouse gas emissions from food systems: building the evidence base. Environmental Research Letters, 16(6), 065007. https://www.fao.org/3/cb5216en/cb5216en.pdf

UN (2020). SDG Partnership Guidebook: A practical guide to building high impact multi-stakeholder partnerships for the SDGs. https://sustainabledevelopment.un.org/content/documents/2698SDG_Partnership_Guidebook_1.01_web.pdf

UN (2022, February 2). The Partnership Platform. Sustainable Development; Department of Economic and Social Affairs, UN https://sdgs.un.org/partnerships

UPU (2017, October 31). Country specific list of prohibited and restricted articles. Universal Postal Union (UPU). https://www.upu.int/UPU/media/upu/files/postalSolutions/programmesAndServices/postalSupplyChain/customs/prohibitedArticles/ListOfProhibitedArticles.pdf

Verdouw, C., Tekinerdogan, B., Beulens, A., & Wolfert, S. (2021). Digital twins in smart farming. Agricultural Systems, 189, 103046. https://doi.org/10.1016/j.agsy.2020.103046

Washington Post (2022, July 22). A drought in Italy's risotto heartland is killing the rice. By Chico Harlan and Stefano Pitrelli https://www.washingtonpost.com/world/2022/07/22/italy-drought-2022-rice-risotto

WBCSD (2022, April). SPHERE: the packaging sustainability framework. World Business Council for Sustainable Development (WBCSD) https://www.wbcsd.org/contentwbc/download/14021/202395/1

WEF (n.d.). SDG https://intelligence.weforum.org/topics/a1G0X0000057N0oUAE and Global issue https://intelligence.weforum.org/topics/a1G680000004Cv2EAE by the World economic forum (WEF).

Wilkinson, M., Dumontier, M., Aalbersberg, I. ... Mons, B. (2016, March 15). The FAIR Guiding Principles for scientific data management and stewardship. Scientific Data, 3(1), 1-9 (160018) https://doi.org/10.1038/sdata.2016.18

## Additional Information

**Competing Interests statement:** There are no competing interests; financial and non-financial.

**Author Contribution statement:** R.P. wrote the manuscript text.